\documentclass[RNAAS]{aastex62}
\usepackage{amsmath}

\begin{document}

\title{An Empirical Mass-Radius Relation for Cool Giant Planets}
\author{Daniel P. Thorngren}
\affil{Department of Physics, University of California, Santa Cruz}
\affil{Institut de Recherche sur les Exoplan\`etes, Universit\'e de Montr\'eal, Canada}
\author{Mark S. Marley}
\affil{NASA Ames Research Center, Moffett Field}
\author{Jonathan J. Fortney}
\affil{Department of Astronomy and Astrophysics, University of California, Santa Cruz}
\maketitle

A number of applications, often related to estimating the yields of direct imaging extrasolar planet searches \citep[e.g.,][]{Stark2019} or estimating planetary radii through Bayesians retrievals on planets with constrained masses \citep[e.g.,][]{Nayak2017}, require the ability to probabilistically predict the radius of a giant planet given its mass \citep[e.g.][]{Wolfgang2016, Chen2017}. However some popular methods currently in use neglect to distinguish between cool giants and inflated hot planets \citep[see][]{Mayorga2018}. Since these planetary populations have very different radii, with the cooler gas giants being smaller, these expressions tend to estimate radii that are too large for cool giants.  Using inappropriate inflated radii when modeling the reflected light signal for these planets (for instance) will yield expected signals that are unrealistically strong. In this work, we will identify a simple, easily applied empirical relationship between the mass and radius of giant planets below 1000 K, which have been shown to be reliably uninflated \citep{Miller2011, Demory2011}.

We downloaded our data from the NASA Exoplanet Archive \citep{Akeson2013} and exoplanets.eu \citep{Schneider2011}, from which we selected the 81 cool giant exoplanets between $15 M_\oplus$ and $12 M_J$ with measured masses and radii.  We fit these with a heirarchical Bayesian model which accounted for uncertainties in the masses and radii as well as the underlying radius dispersion.  For parameters $\vec{\theta}$, this resulted in the following posterior:
\begin{align}
    p(\vec{\theta}, \vec{M}, \vec{R} | R_{obs}, M_{obs}, \vec{\sigma}_{M}, \vec{\sigma}_{R}) \propto
        p(\vec{\theta}) \sum_i
        \mathcal{N} \left(
            R_{obs} | R_i, \sigma_{R,i}
        \right)
        \mathcal{N} \left(
            M_{obs} | M_i, \sigma_{M,i}
        \right)
        \mathcal{N} \left(
            R_i | R(M_i, \vec{\theta}), \sigma(M_i, \vec{\theta})
        \right)
\end{align}
Here $\vec{R}_{obs}, \vec{M}_{obs}, \vec{\sigma}_M, \vec{\sigma}_R$ are the observed radii, masses and their observational uncertainties.  Subscripts indicate the $i^{th}$ element of the vector.  $M_i$ and $R_i$ are nuisance parameters for the actual mass and radius of the planet, which we marginalize out.  Finally,  $R(M, \vec{\theta})$ and $\sigma(M, \vec{\theta})$ are the mass-radius relationship and its dispersion.  We considered a wide variety of functional forms for these with varying numbers of parameters and uninformative priors.  These were evaluated for their predictive power using the AIC \citep[Akaike Information Criterion][]{Akaike1974}, a commonly used model selection criterion \citep[see][]{Gelman2014}.  As more planets are discovered, more complex models will be favored.  Based on this, we selected as our preferred model a quadratic relation between radius and log mass with power-law dispersion:
\begin{align}\label{theFit}
R(M) &= c_0 + c_1 \log(M) + c_2 \log(M)^2 \pm \sigma_0 M^\alpha
\end{align}

\begin{figure}[t]
    \centering
    \includegraphics[width=\textwidth]{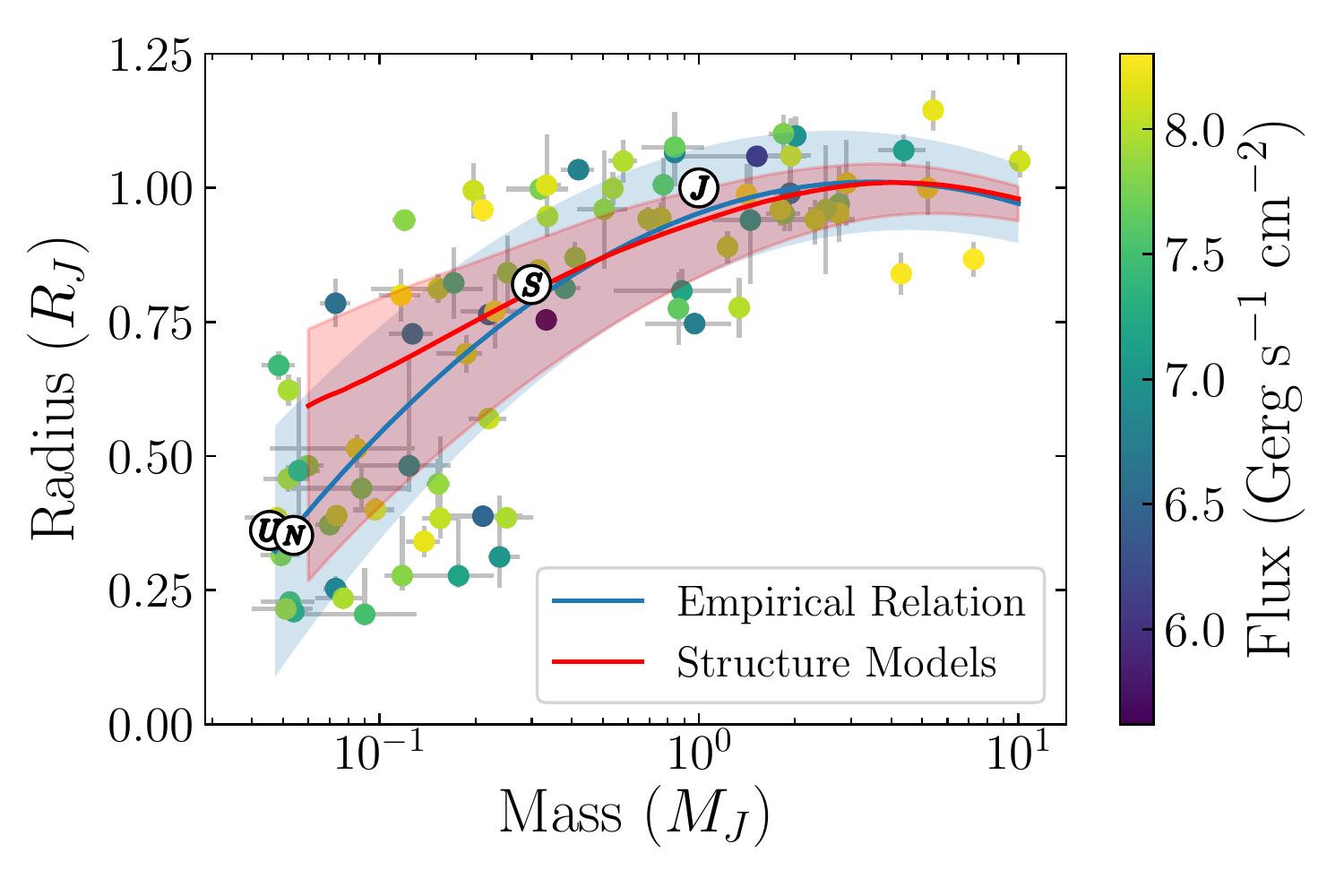}
    \caption{The masses and radii of 81 observed cool giant planets with $1 \sigma$ errorbars, along with Jupiter, Saturn, Uranus and Neptune.  The $\log_{10}$ of the incident flux is indicated by the color, to show that it does not affect the radius.  The fit and $1 \sigma$ dispersion from Eq. \ref{theFit} and the posterior mean parameters are shown in blue.  The relation from the structure models of \citep{Thorngren2016} is shown in red, where the dispersion is from the uncertainty in the mass-metallicity relation.}
    \label{theFigure}
\end{figure}

Our MCMC retrieval yields the parameter estimates $c_0 = .96 \pm .02$, $c_1 = .21 \pm .03$, $c_2 = -.20 \pm .042$, $\sigma_0 = .12 \pm .015$, and $\alpha = -.215 \pm .069$.  The posterior was approximately multivariate normal with only small correlations.  Figure \ref{theFigure} shows the resulting curve and dispersion for the posterior mean, compared against the observed planets and the structure model results of \cite{Thorngren2016} (the radius dispersion comes from the dispersion in the mass-metallicity relation).  The two lines align closely at higher masses, with a modest disagreement at lower masses, primarily caused by the strong left-skew in the structure model predictions.  Jupiter, Saturn, Uranus, and Neptune lie within the 1 $\sigma$ contours of both predictions.

Using this formula and the parameter values provided, one may derive an estimated radius and predictive uncertainty for cool giant planets with known masses.  Note that this formula should only be applied to planets with $T_{eq} < 1000$ K with masses between $15 M_\oplus$ and $12 M_J$; because this is a simple empirical fit, extrapolation is inappropriate.  Within this range, however, our predictions will be more accurate than most other methods, which do not account for $T_{eq}$.  Also, very young planets (up to hundreds of megayears) may retain enough formation heat to be larger than our relation.  We anticipate that this relation will be useful to the community for both the planning and interpretation of giant planet observations.

\bibliography{bibliography}
\end{document}